\documentclass[preprint]{elsarticle}

\usepackage{graphicx,subfigure}
\usepackage{epstopdf}
\usepackage{amsmath, amssymb}
\usepackage{rotating}
\usepackage{morefloats}
\usepackage{comment}

\begin{document}
\title{Physical approach to price momentum and its application to momentum strategy}
\author[sbp]{Jaehyung Choi\corref{cor1}}
\ead{jaehyung.choi@sunysb.edu}

\cortext[cor1]{Correspondence address: Department of Physics and Astronomy, SUNY at Stony Brook, NY 11794-3800, USA. Fax:+1-631-632-8176.}
\address[sbp]{Department of Physics and Astronomy\\SUNY Stony Brook, NY 11794-3800, USA}

\begin{abstract}
	We introduce various quantitative and mathematical definitions for price momentum of financial instruments. The price momentum is quantified with velocity and mass concepts originated from the momentum in physics. By using the physical momentum of price as a selection criterion, the weekly contrarian strategies are implemented in South Korea KOSPI 200 and US S\&P 500 universes. The alternative strategies constructed by the physical momentum achieve the better expected returns and reward--risk measures than those of the traditional contrarian strategy in weekly scale. The portfolio performance is not understood by the Fama--French three-factor model.
\end{abstract}
\maketitle

\section{Introduction}
	Searching for the existence of arbitrage is an important task in finance. In the case of the statistical arbitrages, regardless of their origins such as market microstructure, firm-specific news/events, and macroeconomic factors, it is possible to exploit arbitrage opportunities via trading strategies in order to take consistent profits. Among such kinds of the statistical arbitrage chances, they are also called as market anomalies, if their origins are not well-explained or understood quantitatively and qualitatively \cite{Lo:2001p5538, Singal:2006p5475}. To academic researchers in finance, it is very useful for testing the robustness of the efficient market hypothesis \cite{Fama:1965p4897,Samuelson:1965p4904} and no-arbitrage theorem. Although they had played the keystone roles in asset pricing theory and general finance, their statuses recently have been changed as alternative theories that intrinsically allow the pricing anomalies have appeared in financial markets, as instances, the adaptive market hypothesis \cite{Farmer:1999,Lo:2004,Lo:2005} and behavioral finance \cite{Kahneman:1982,Shleifer:1995,Shleifer:2000,Kahneman:2000}. Hunting for the systematic arbitrage opportunities is also crucial to market practitioners such as traders and portfolio managers on Wall street because it is the core of money-making process that is their most important role.
	
	Among these market anomalies, price momentum has been the most well-known example to both groups. Since Jegadeesh and Titman's seminal paper \cite{Jegadeesh:1993p200}, it has been reported that the prices of financial instruments exhibit the momentum effect that the future price movement tends to keep the same direction along which it has moved during a given past period. It is also realized that the momentum strategy, a long-short portfolio based on the momentum effect, has been a profitable trading strategy in the stock markets of numerous developed and emerging countries during a few decades even after its discovery \cite{Rouwenhosrt:1998,Rouwenhosrt:1999}. In addition to the existence in equity markets, the momentum effect large enough to implement as the trading strategy is also found in other asset classes such as foreign currency exchange \cite{Okunev:2003}, bond \cite{Asness:2008}, futures \cite{Asness:2008,Moskowitz:2010}, and commodities markets \cite{Erb:2006}.
	
	In spite of its success in profitability over diverse asset classes and markets, its origin has not been fully understood in the frame of traditional mainstream finance. This is why the momentum effect is one of the most famous market anomalies. Attempts to explain the momentum effect with factor analysis have failed \cite{Fama:1996} and the reason the momentum effect has persisted over decades still remains mysterious. The Fama--French three factor model is able to explain small portions of the momentum return \cite{Fama:1996}. The lead-lag effect or auto-/cross-sectional correlation between equities are one of the possible answers to the momentum effect \cite{Lo:1990p883,Lewellen:2002}. The sector momentum is another partial interpretation on the anomaly \cite{Moskowitz:1999p4294}. Additionally, the behavioral aspects of investors such as collective response to financial news and events have broadened the landscape of understanding on the momentum effect \cite{Hong:1999p4506, Terence:1998p4385, Daniel:1998p4514, Barberis:1998p307}. Transaction cost is also considered a factor which caused the momentum effect \cite{Lesmond:2004}. Unfortunately, none of these explanations are capable of providing the entire framework for explaining why the momentum of price dynamics exists in many financial markets.
	
	Not only demystification on the origins of the price momentum, pursuit on the profitability and implementability of the momentum effect in financial markets also have been interesting to academia and practitioners. For example, although several studies \cite{Koh:1997, Ahn:2004, Chae:2009, Asness:2011, Liu:2011} found that the momentum strategies in some Asian markets such as Japanese stock market are not profitable, Asness et al. \cite{Asness:2008} discovered that the momentum strategy in Japan becomes lucrative, when it is combined with other negatively correlated strategies such as value investment. Not limited to several stock markets, the hybrid portfolio of value and momentum also outperforms each of the value and momentum portfolios across the asset classes. Their study paid attention to the implementation of the momentum strategy combined with fundamental value investment factors such as book-market (BM) ratio\footnote{It is also related to price--book (PB) ratio inversely. Many literature on momentum mostly use BM ratio as a momentum-driven factor and PB ratio also known as PBR is frequently mentioned in fundamental analysis of stocks.} which also has been used to unveil the origins of the momentum effect in Fama--French three factor analysis. In other words, their work can be understood as the construction of the hybrid portfolio to increase the profitability and stability of the portfolios based on the momentum strategy. Moreover, the selection criteria for the hybrid portfolio are considered as the multiple factors related to the momentum returns whether they are positively correlated or negatively correlated. Academically, this observation imposes the important meaning in the sense that these multiple filters can explain their contributions to the momentum returns. In practical viewpoint, it is obviously the procedure for generating trading profits in the markets.
	
	Another method for improving the profitability of the momentum strategy is introducing various selection rules for the construction of the momentum portfolio. First of all, simple variations in the original momentum selection rule can be made. Moskowitz et al. \cite{Moskowitz:2010} suggested new trading strategies based on time series momentum which constructs the momentum portfolios by time series regression theory. It is not simply from cumulative return during a lookback period as a sorting variable but from an autoregressive model of order one which can forecast the future returns under given conditions such as the past returns and volatilities. The predicted return is used as the selection criterion for the time series momentum strategy. The time series momentum portfolio performs very well even during market crisis. It also shares the common component which drives the momentum return with the cross-sectional momentum strategy across many asset classes. This fact imposes that the momentum strategy is improved by the modified cumulative return criterion and there is a possibility to find the better momentum strategies in performance and risk. 
	
	Besides only considering the cumulative return, the introduction of alternative proxies for the portfolio selection rules has been also worth getting attention. George and Hwang \cite{George:2004} used 52-week high price\footnote{The 52-week high price is the highest price during last 52 weeks, i.e. 1 year.} as a selection criterion and the momentum portfolio based on the 52-week high price generated stronger returns. Additionally, the tests on the momentum portfolios, which are doubly-sorted by the cumulative return or sector momentum and the 52-week high price, exhibit the superiority of the 52-week high price criterion. The factor analysis also shows that the return from the 52-week high price factor is not only stronger than the traditional and sector momentum factors but also statistically more significant and important in the momentum return modeling. The dominance of the 52-week high momentum criterion is also observed in the various international stock markets \cite{Liu:2011}.
	
	Reward-risk measures are also able to serve as the ranking criteria. Rachev et al. \cite{Rachev:2007p616} used the reward--risk measures as the sorting criteria for their momentum portfolios instead of the cumulative return over the estimation period. In their work, Value-at-Risk (VaR), Sharpe ratio, R-ratio, and STARR were used as alternative ranking rules. In the S\&P 500 universe from 1996 to 2003, their momentum portfolios constructed by the reward--risk measures provided the better risk-adjusted returns than the traditional momentum strategy. In addition to that, the new momentum portfolios had lower tail indexes for winner and loser baskets. In other words, these momentum strategies based on the reward--risk measures obtained the better risk-adjusted returns with acceptance of the lower tail risk.
	
	Back to physics, the momentum in price dynamics of a financial instrument is also an intriguing phenomenon because the persistent price dynamics and its reversion can be understood in terms of inertia and force. The selection rules of the momentum strategy is directly related to the ways of how to define and measure ``physical'' momentum in price dynamics of the instrument. When the instrument is considered as a particle in a one-dimensional space, the price momentum is also calculated if mass and velocity are defined. Since the momentum effect exists, it can be concluded that price of an equity has inertia that makes the price keep their direction of movements until external forces are exerted. In this analogy, the external force corresponds to any exogenous market events and information such as good/bad news, changes in psychology and macroeconomic situation, and imbalance in supply and demand. This idea is also able to explain why the cumulative return based momentum strategy generates the positive expected returns. However, it has been not much attractive to physicists yet. Most of the econophysics community has not been interested in trading strategy and portfolio management so far.
	
	Recently, Choi \cite{Choi:2011} suggested that a trading strategy can be considered as being in the spontaneous symmetry breaking phase of arbitrage dynamics. In his work, the return dynamics had a parity in the return which can be broken by choosing the ground state. When a control parameter is smaller than a critical value, the strategy is in the arbitrage phase and we expect the non-zero expected return which is not permitted in the efficient market hypothesis. Random fluctuation around the non-zero value makes variance of the strategy return and the risk of loss still exists. The important caveats were not only that the arbitrage strategy can theoretically generate the non-zero expected returns emergent from the symmetry breaking but also that the idea is empirically meaningful when it is applied to real trading strategies. For the simple back-testing, the control parameter, which triggers phase transition, was estimated from an autocorrelation coefficient of the strategy return time series. If the strategy is expected to be in the arbitrage phase, the strategy is exploited and if not, the execution is stopped. As an empirical test, the weekly contrarian strategy was executed based on the scheme using the symmetry breaking arbitrage. The contrarian strategy with the scheme had the better expected return and Sharpe ratio than the traditional contrarian strategy.
	
	In this paper, we introduce various definitions for the physical momentum of equity price. Based on those definitions, the equity price momentum can be quantified from real historical data in the South Korean KOSPI 200 and S\&500 universes. After computing the physical momentum, the implementation of the contrarian strategies based on the candidates for the price momentum increases the validity of our approach for measuring the physical momentum in equity price. Empirically, these new candidates for the selection criteria which originated from the physical momentum idea provide the better returns and Sharpe ratios than the original criterion, i.e. the cumulative return. The structure of this paper is the following. In the next section, the definition of velocity in equity price space and possible candidates for financial mass are introduced and then the price momentum is defined with the financial velocity and mass. In section \ref{sec_phys_data}, we specify the datasets used for our analysis. In section \ref{sec_phys_result}, results for the physical momentum strategies are given. The Fama--French three-factor analysis is given in section \ref{sec_phys_factor_analysis}. In section \ref{sec_phys_conclusion}, we conclude the paper.
		
\section{Theoretical background}
\label{sec_phys_background}
	When a one-dimensional space for price of a financial instrument is introduced, it is possible to consider that the price is in motion on the positive half-line. Although the negative price is conceptually proposed by Sornette \cite{Sornette:2000p4733}, the negative price of the instrument is not allowed in practice.\footnote{Sornette not only pointed out that the negative equity price is introduced only for symmetry breaking but also explained why the negative price is not observed under real situations using dividend payment as an external field in symmetry breaking.} The price dynamics of the financial instruments are now changed to a one-dimensional particle problem in physics. To extend the space to the entire line, the log price is mapped to the position $x(t)$ in the space by
	\begin{eqnarray}
		x(t)=\log{S(t)}\nonumber
	\end{eqnarray}
	where $S(t)$ is the price of the instrument. This transformation is not new to physicists because Baaquie \cite{Baaquie:1997,Baaquie:2004} already introduced the same transformation to derive path integral approach to option pricing theory. It was used in order to find the relation between the Black-Scholes equation and the Schr\"odinger equation. With this re-parametrization, an option pricing problem was transformed to a one-dimensional potential wall problem in quantum mechanics. However, it was not for introducing the physical momentum concept mentioned above. With the log return, $x(t)$ covers the whole line from the negative to positive infinity. In addition to the physical intuition, the log price has some advantages in finance. First of all, it is much simpler to calculate the log return from the log price because the difference of two log prices is the log return. Contrasting to the log return, the raw return is more complicated to compute than the log return. Secondly, one of the basic assumptions in mathematical finance is that the returns of financial instruments are log-normally distributed and we can handle normally-distributed log returns.
	
	With the advantages of the log price described above, it is natural to introduce a concept of velocity into the one-dimensional log price space. In the case of the log price, the log return $R(t)$ per time scale is expressed in $x(t)$ by
	\begin{eqnarray}
		R(t)&=&\frac{\log{S(t)}-\log{S(t-\Delta t)}}{\Delta t}\nonumber\\
		&=&\frac{x(t)-x(t-\Delta t)}{t-(t-\Delta t)}\nonumber\\
		&=&\frac{\Delta x(t)}{\Delta t}.\nonumber
	\end{eqnarray}
	In the limit of infinitesimal time interval ($\Delta t \to 0$), the log return becomes
	\begin{eqnarray}
		R(t)=\frac{d x(t)}{dt}=v(t)\nonumber
	\end{eqnarray}
	where $v(t)$ is the velocity of the instrument in the log price space, $x(t)$. When the mapping between the log price and position in the one-dimensional space is introduced, it imposes the relation between the log return and velocity. Although this relation works only in the limit of $\Delta t \to 0$, it can be used as the approximation in the discrete time limit if the length of the whole time series is long enough to make the time interval relatively shorter.
	
	The cumulative return $r(t)$ is expressed in $v(t)$ by
	\begin{eqnarray}
		r(t)&=&\frac{S(t)-S(t-\Delta t)}{S(t-\Delta t)}=\exp{(R(t))}-1\nonumber\\
		&=&v(t)\big(1+\frac{1}{2}v(t)+\cdots+\frac{1}{n!}(v(t))^{n-1}+\cdots\big).\nonumber
	\end{eqnarray}
	Since the log return is usually small such as $|v(t)|\ll1$ in real data, higher-order terms in $v(t)$ can be treated as higher-order corrections on $r(t)$ and it is possible to ignore the higher-order corrections if $|v(t)|\ll1$. In this sense, the cumulative return can be approximated to $v(t)$. However this relation is broken in the cases of heavy tail risks caused by financial crisis or firm-specific events such as bankruptcy, merger and acquisition, and good/bad earning reports of the company. Since $|v(t)|$ in these events can be comparable to one or greater than one, the higher-order perturbations should be considered.
	
	Based on the correspondence, the concept of the price momentum can be quantified by using the classical momentum in physics with
	\begin{eqnarray}
		p=mv\nonumber
	\end{eqnarray}
	where $m$ has the same role to physical mass. In particular, when velocity is given in the log return, the contribution of the mass to the price momentum can be expressed in the following way,
	\begin{eqnarray}
		p&=&m\log{(1+r)}\nonumber\\
		&=&\log{(1+r)^m}.\nonumber
	\end{eqnarray}
	The financial mass $m$ plays a role of amplifying the price change as the mass becomes larger. This amplification is understood as filtering of market information on price. Consideration on the volume can catch more market information. For example, large transaction volumes at the peak or trough could impose the change in the trend from the viewpoint of technical analysis.  Some instruments are heavily influenced by the investors' psychology and other market factors but others are not. In this sense, the mass can act a role of the filter which is unique to each instrument and encodes the instrument-specific characteristics. This interpretation is also well-matched to the physical analogy that mass is a physical constant which is unique to each particle. The original ranking criterion in the traditional momentum strategy is a special case of this physical momentum definition. In the cumulative return momentum strategy, it is assumed that each of equities has the identical mass, $m=1$. However, the identical mass assumption seems not to be reasonable because each equity has distinct properties and shows inherent price evolutions. In order to capture these heterogeneities between the characteristics of each equity, the departure from the identical financial mass for all equities is more natural and the introduction of the financial mass concept to the momentum strategy looks plausible. Although the physical momentum concept can be applied to other asset classes, we focus only on the equities in this paper.
	
	As described in the previous paragraph, the financial mass can convey the instrument-specific information. However, it is obvious that all kinds of information cannot work as the candidates for the mass because it should be well-matched to intrinsic properties of physical mass. In this sense, liquidity is a good candidate for the financial mass. Its importance in finance is already revealed in many financial literature in terms of bid-ask spread, volume, and fractional turnover rate \cite{Datar:1998, Amihud:1986, Hu:1997, Lee:2000,Amihud:2002,Kim:2012}. In particular, Datar et al.\cite{Datar:1998} reported that past turnover rate is negatively correlated to future return. With the same size of the momentum, the larger turnover rate brings the poorer future return i.e. illiquid stocks exhibit higher average returns. Even after controlling other factors such as firm-size, beta\footnote{The beta in finance is the correlation between the stock return and benchmark return scaled by the market variance.}, and BM ratio, the past turnover rate has the significant negative correlation with the future return. It is possible to understand that the trading volume incorporates integrated opinions of investors and makes the price approach to the equilibrium asymptotically. In the viewpoint of information, trading can be understood as the exchange of information between investors with inhomogeneous information. More transactions occur, more information is widely disseminated over the whole market and the price change becomes more meaningful. Lee and Swaminathan \cite{Lee:2000} also provided the similar result that stocks with low past trading volumes tend to have high future returns. Additionally, the study found that the momentum strategy among high volume stocks is more profitable. The similar result is obtained in the South Korean market \cite{Kim:2012}. 
	
	The possible mass candidates, which are also well-matched to the analogy of physical mass, are volume, total transaction value, and inverse of volatility. If the trading volume is larger, the price movement can be considered as the more meaningful signal because the higher volume increases the market efficiency. The amount of the volume is proportional to mass $m$. As mentioned in the previous paragraph, the relation between the trading volume and asset return is already studied in finance \cite{Datar:1998, Hu:1997, Lee:2000}. Instead of the raw volume, we need to normalize the daily volume with the total number of outstanding shares and this normalized value is also known as a turnover rate. The reason of this normalization is that some equities intrinsically have the larger trading volumes than others because the total number of shares enlisted in the markets are much larger than other equities or because they get more investors' attention which causes more frequent trades between investors. The share turnover rate, i.e. trading volume over total outstanding shares is expressed in $\upsilon$ in the paper. 
	
	Similar to the volume, the daily transaction value in cash can be used as the financial mass. If an equity on a certain day has the larger transaction value, investors trade the equity frequently and the price change has more significant meanings. Additionally, the transaction value contains more information than the volume. For examples, even though two equities record the same daily volume and daily return on a given day, the higher priced equity exhibits the larger trading value if two prices are different. The more important meaning is that even though market information such as close price, volume, return, and price band are identical, the trading value in cash can be different. As an instance, when one equity is traded more near the lowest price of the daily band but the other is traded mainly around the daily highest price region, the total transaction values of two equities are obviously different. It also needs to be normalized because each equity price is different. The normalization of dividing total transaction value by market capitalization is expressed in $\tau$ in the paper. 
	
	The return volatility $\sigma$ is inversely proportional to the financial mass $m$. If the volatility of a certain equity in a given period is larger, the equity price is easy to fluctuate much severely than other equities with the smaller volatilities. This corresponds to the situation in physics where a lighter object can move more easily than a heavy object under the same force. So if it is heavy in the sense of the volatility, the asset price with larger mass is under the smaller volatility. This definition of the financial mass is also matched with the analogy used in Baaquie's works \cite{Baaquie:1997, Baaquie:2004}. In his works, the Black-Scholes equation was transformed into Hamiltonian of a particle under the potential which specifies the option. The mass of a particle in the Hamiltonian was exactly same to the inverse of the return volatility. Since the volatility is also interesting to economists and financiers, there are series of the literature covering the link between volatility and return \cite{Glosten:1993, French:1987}.
	
	With the fractional volume and fractional transaction value as the proxies for the mass, it is possible to define two categories of the physical momentum,
	\begin{eqnarray}
		p^{(1)}_{t,k}(m,v)=\sum_{i=0}^{k-1} m_{t-i} v_{t-i}\nonumber
	\end{eqnarray}
	and
	\begin{eqnarray}
		p^{(2)}_{t,k}(m,v)=\frac{\sum_{i=0}^{k-1} m_{t-i} v_{t-i}}{\sum_{i=0}^{k-1} m_{t-i}}\nonumber
	\end{eqnarray}
	over the period of the window size $k$. The latter one is the reminiscent of the center-of-mass momentum in physics and the similar concept is used as the embedded capital gain in Grinblatt and Han \cite{Grinblatt:2002}. Since two different categories for the momentum calculation, two for return, and two for mass are available, there are eight different momentum definitions for an equity. 
	
	It is easily found that the cumulative return can be expressed in $p^{(1)}$ by
	\begin{eqnarray}
		r_{t,k}&=&\exp{(\sum_{i=0}^{k-1} R_{t-i})}-1=\exp{(p^{(1)}_{t,k}(1,R))}-1\nonumber\\
		&\approx& p^{(1)}_{t,k}(1,R)+\mathcal{O}\Big(\big(p^{(1)}_{t,k}(1,R)\big)^2\Big)\nonumber
	\end{eqnarray}
	and this shows that the traditional momentum in finance is a special case of the physical momentum. In this sense, let us call $r_{t,k}=p^{(0)}_{t,k}$. In addition to that, since exponential function and log function are strictly increasing functions, the mapping between $p^{(0)}_{t,k}$ and $p^{(1)}_{t,k}(1,R)$ is one-to-one.
	
	Since the return volatility over the period contains more practical meanings than the sum of daily volatilities during the period, the third class of the physical momentum is defined by
	\begin{eqnarray}
		p^{(3)}_{t,k}(m,v)=\bar{v}_{t,k}/\sigma_{t,k}\nonumber
	\end{eqnarray}
	where $\bar{v}_{t,k}$ is the average velocity at time $t$ during the past $k$ periods. There are also two different definitions for $p^{(3)}_{t,k}$ computed from the normal return and log return. This is closely related to the Sharpe ratio, $SR$,
	\begin{eqnarray}
		SR=\frac{\mu(r-r_f)}{\sigma(r-r_f)}\nonumber
	\end{eqnarray}
	where $r_f$ is the risk-free rate. If the risk-free rate is small and ignorable, $p^{(3)}_{t,k}$ approaches the Sharpe ratio. The momentum strategy with this ranking criterion is the reminiscent of the Sharpe ratio based momentum strategy by Rachev et al. \cite{Rachev:2007p616}. Similar to the Sharpe ratio, $p^{(3)}_{t,k}$ can be related to the information ratio that uses excessive returns over the benchmark instead of the risk-free rate in the definition. However, we neither consider the risk-free rate nor the benchmark return as a reference point of the portfolio returns in this paper.
	
	With $p^{(1)}_{t,k}$, $p^{(2)}_{t,k}$, and $p^{(3)}_{t,k}$, total eleven different definitions of physical momentum including the traditional cumulative return are the possible candidates for the physical equity momentum. Each of them is originated from the physical and financial foundations. Additionally, they are relatively easier to quantify than other risk measures used in Rachev's work \cite{Rachev:2007p616}. Although it is possible to consider more complicated functions of other market data for the price momentum, it is beyond the scope of this paper.

\section{Application to real data}
\label{sec_phys_data}
\subsection{Dataset}
\subsubsection{South Korea equity markets: KOSPI 200}
	The market data and component-change log of the KOSPI 200 universe are downloaded from Korea Exchange. The covered time horizon starts January 2003 and ends in December 2012.

\subsubsection{U.S. equity markets: S\&P 500}
	The daily market information and roaster for S\&P 500 components are collected from Bloomberg. The time window is identical to the KOSPI 200.
\subsection{Momentum/Contrarian strategy}
	The comparison of the performances between the traditional momentum strategy and physical momentum strategy is able to test the validity of the physical momentum definition. Instead of the traditional momentum strategy that uses the raw return during the lookback period as a ranking criterion, we can construct the momentum portfolio ranked by the various definitions of the physical momentum. After finding the performance, each of the momentum strategies from the various momentum criteria is compared with others in order to measure the validity of a given momentum definition. Details on the momentum strategy will follow.

	The most important variables of the momentum strategy are the length of the lookback (or estimation) period $J$, the length of the holding period $K$, and the sorting criterion $\psi$. The traditional momentum strategy uses the cumulative return during the lookback period as a ranking criterion, i.e. a triplet of the traditional momentum strategy is $(J, K, \psi=p^{(0)})$ \cite{Jegadeesh:1993p200}. On the reference day ($t=0$), the cumulative returns of all instruments in the market universe during the periods from $t=-J$ to $t=-1$ are calculated. After sorting the instruments in the order of the ascending criterion, numbers of ranking groups are constructed and each of the ranking groups has the same number of the instruments. As an instance, if there are 200 equities and we consider 10 groups, each of sorted ranking groups has 20 equities as group constituents. Following the convention of Jegadeesh and Titman \cite{Jegadeesh:1993p200}, the loser group that has the worst performers in the market is named as R1 and the winner group with the best performers is the last one, R10. And then the momentum portfolio is constructed by buying the winners and short-selling the losers with the same size of positions in cash in order to make the composite portfolio dollar-neutral. For the winner and loser portfolios, each group member is equally weighted in the group in which it is. The constructed momentum portfolio is held until the end of the holding period ($t=K$). On the last day of the holding period ($t=K$), the momentum portfolio is liquidated by selling the winner group off and buying the loser group back.
	
	On the first day of each unit period, the momentum portfolio is constructed as explained in the previous paragraph. For example, a weekly momentum portfolio is selected on every Monday unless it is not a holiday. Monthly portfolios are formulated on the first business day in every month. For multiple-period holding strategies, there exists overlapping period between two strategies constructed by the same criterion at the different reference dates. The reasons of this construction are the following. First of all, the momentum return from this construction is not dependent on the starting point of the strategy formation. For example, when we implement the 12-month lookback and 12-month holding momentum strategy, construction of the portfolio occurs at the beginning of each year. Since the return results are always interfered by the seasonal effects such as January effect or others related to business cycle and taxation, it is difficult to discern the momentum effect from the seasonal effects. Second, the portfolios from overlapped periods can generate larger number of return samples to fortify the statistical significance. Since the dataset here only has twelve years of historical data comparing with other studies which use much longer time periods as datasets, its statistical significance could be lowered by the small size of our samples if we use the non-overlapped portfolios. Third, Jegadeesh and Titman \cite{Jegadeesh:1993p200} reported that there were not big differences between the returns by the overlapped and non-overlapped portfolios. Finally, the portfolio construction here can be considered as diversification which helps to mitigate the large fluctuation of returns in the momentum portfolio. For example, in the case of 12-month holding strategies, we possess 12 different portfolios at a given moment and it is definitely a diversification of the portfolio. Based on these reasons, it is more sensible that the overlapping portfolios are used in our case.
	
	When we buy the winner and loser portfolios which provide expected returns for those groups of $r_W$ and $r_L$ respectively, the return by the momentum portfolio $r_{\Pi}$ is simply $r_{\Pi}=r_W-r_L$ because we short-sell the losers in the portfolio. When we implement the trading strategy in the real financial markets, a transaction cost including brokerage commission and tax is always important because they actually erode the trading profits. The implemented momentum return or transaction-cost-adjusted return $r_I$ is 
	\begin{eqnarray}
		r_I&=&r_{\Pi}-c\nonumber\\
		&=&(r_W-r_L)-(c_W+c_L)\nonumber
	\end{eqnarray}
	where $c_W$ and $c_L$ are the transaction costs for the winner and loser groups, respectively. In general, $c_L$ is greater than $c_W$ because the short-selling is usually much more difficult than buying. Since the transaction cost is a one-time charge, its effect on the implemented return per unit period becomes smaller as the holding period is lengthened. 
	
	When the expected return of the momentum portfolio for a given $(J, K, \psi)$ strategy is negative, the strategy can become profitable by simply switching to the contrarian strategy $(J, K, \psi^\dagger)$ that buys the past loser group and short-sells the past winner group, exactly the opposite position to the momentum portfolio. Contrasting to the momentum strategy following the price trend, the contrarian strategy is based on the belief that there is the reversion of price dynamics. If equities have performed well during the past few periods, investors try to sell those stocks to put the profits into their pockets. The investors who bought those equities long time ago are able to make large enough profits even when the price recently has gone slightly downward. However, buyers who recently purchased the equities might not have enough margins yet from their inventories and want not to lose money from the current downward movement because of risk aversion. The only option those investors can take is just selling their holdings off. This herding behavior makes the reversion of price and it is probable to make profits from short-selling if a smarter investor knows when the timing would be. For the opposite case, it is also possible to buy the past losers to get advantage of using the herding because the losers are temporarily undershot by investors' massive selling force and the equities tend to recover their intrinsic values. On the way of price recovery, the short-sellers need to buy back what they sold in the past in order to protect their accounts and the serial buy-back can boost the price dynamics to the upward direction which also causes the consequential massive buy-backs by other short-sellers. How much the initial anomaly can be amplified is modeled in Shleifer and Vishny \cite{Shleifer:1995}.
	
	The momentum and contrarian strategies look contradictory to each other but they have only the different time horizons in which each of strategies works well. Usually, in three to twelve months scale, the equity follows the trends \cite{Jegadeesh:1993p200} but the reversal effect is dominant at the longer and shorter scales than the monthly scale \cite{Lo:1990p883,DeBondt:1985p4232}. For the contrarian strategy, the portfolio return $r_{\Pi^{\dagger}}$ is given by 
	\begin{eqnarray}
		r_{\Pi^{\dagger}}=r_L-r_W=-r_{\Pi}\nonumber.
	\end{eqnarray}
	The transaction cost adjusted return $r_I$ for the contrarian strategy is 
	\begin{eqnarray}
		r_I&=&r_{\Pi^{\dagger}}-c\nonumber\\
		&=&(r_L-r_W)-(c_W+c_L)\nonumber.
	\end{eqnarray}
	When the implementability of a given strategy in the real markets is the main concern, we need to focus on whether or not it is possible to take actual profits from the strategy among the momentum and contrarian strategies. In this sense, the profitability of the strategy with absolute (implemented) return $\tilde{r}_I$ can be measured by 
	\begin{eqnarray}
		\tilde{r}_I=|r_W-r_L|-(c_W+c_L)\nonumber
	\end{eqnarray}
	and tells whether the potential trading profit can exceed the barrier of the transaction cost. The actual positive return from the momentum/contrarian trading strategies can be taken into the pocket when $\tilde{r}_I$ is positive. However, the transaction cost is not considered because the outperformance/underperformance with respect to the benchmark strategy is our main concern.
	
	As mentioned above, the method for measuring the price momentum is the momentum strategy with the physical momentum as a ranking criterion. There are total eleven types of candidates for physical momentum including the original cumulative return momentum. On the reference day $(t=0)$, each physical momentum for equities over the estimation period of 6 weeks is calculated and used for sorting the equities. The ranking for each criterion constructs the momentum portfolios. After holding the portfolio during 6 weeks, it is liquidated to get the momentum profit. The positive implemented returns and Sharpe ratios from the implemented return exhibit the robustness of the physical momentum strategies. If their returns beat that of the traditional momentum strategy, it is obvious that the physical momentum definition really has a merit to introduce and there is a practical reason to use the momentum strategies based on the physical momentum as an arbitrage strategy.
		
	For the lookback period, some stocks which do not have enough trading dates are ignored from the analysis. In general, this case happens to companies which are enlisted to the market universe amid of the lookback period. If an equity is traded on only one day during the estimation period, it is neglected from our consideration for the momentum universes because it is impossible to calculate the standard deviation for $p^{(3)}$-type momentum for these stocks. Since all possible candidates for the physical momentum need to be compared with other criteria over the same sample, it is obvious not to consider these equities with only one trading day in the estimation period. The companies delisted amid of the holding periods do not cause the same problem because only the lookback return is important in sorting the equities and constructing the momentum portfolios. In this case, the returns for the delisted companies are calculated from the prices on the first and last trading days in the holding periods.
	
\section{Results}
\label{sec_phys_result}
\subsection{South Korea equity market: KOSPI 200}
\begin{figure}[h!]
	\begin{center}
		\includegraphics[width=8cm]{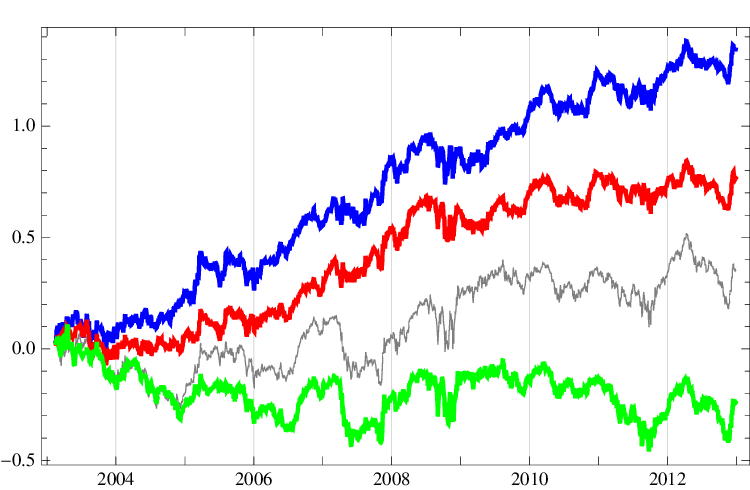}
	\end{center}
		\caption{Cumulative returns for the traditional contrarian (gray), $p^{(1)}(\upsilon,R)$ (blue), $p^{(2)}(\upsilon,R)$ (red), and $p^{(3)}(1/\sigma,R)$ (green) in South Korea KOSPI 200. (For interpretation of the references to color in this figure legend, the reader is referred to the web version of this article.)}
		\label{grp_acc_return_phys_momentum_weekly_6_6_contrarian_kr_kp200}
\end{figure}

	In Table \ref{tbl_daily_summary_stat_risk_phys_momentum_weekly_6_6_contrarian_kr_kp200}, all weekly contrarian strategies based on physical momentum except for the $p^{(3)}$ criteria outperform the traditional contrarian strategy. The $p^{(1)}(\upsilon,R)$ portfolio is the best contrarian strategy with the weekly return of 0.261\% under the volatility of 2.444\% while the benchmark contrarian strategy obtains weekly 0.069\% with the standard deviation of 2.846\%. The performance of other $p^{(1)}$ portfolios is as good as the $p^{(1)}(\upsilon,R)$ portfolio and is better than the $p^{(0)}$ criterion. Although the $p^{(2)}$ strategies are slightly worse than the $p^{(1)}$ cases, the performance is much better and less volatile than the mean-reversion strategy. Although all $p^{(3)}$ portfolios exhibit smaller standard deviations, the average returns are all negative. With a few exceptions, the skewness levels of the alternative weekly contrarian strategies are higher and the kurtosis is lower than those of the cumulative return strategy. The historical performance of the portfolios is given in Fig. \ref{grp_acc_return_phys_momentum_weekly_6_6_contrarian_kr_kp200}.
	\begin{sidewaystable}
\begin{center}
\caption{Summary statistics and risk measures of weekly 6/6 contrarian portfolios in South Korea KOSPI 200}
\scriptsize
\begin{tabular}{l l r r r r r r r r r}
\hline
Criterion & Portfolio & \multicolumn{5}{l}{Summary statistics}  & \multicolumn{4}{l}{Risk measures}\\ \cline{3-7} \cline{8-11} 
 & & Mean & Std. Dev. & Skewness & Kurtosis & Fin. Wealth & Sharpe & $\textrm{VaR}_{95\%}$ & $\textrm{CVaR}_{95\%}$ & MDD \\ 
\hline
$p^{(0)}$& Winner (W) &0.1977&3.8044&-1.2951&6.7842&1.0181&0.0689&1.7492&2.5255&67.27\\
 & Loser (L) &0.2662&4.3629&-1.1757&7.6803&1.3708&0.0818&1.6779&2.5040&63.00\\
 & L -- W &0.0685&2.8461&0.1907&1.8784&0.3527&0.0149&1.6930&2.4344&33.40\\
\\[-2ex] 
$p^{(1)}(\upsilon,r)$& Winner (W) &0.0365&4.3618&-1.2676&6.7068&0.1880&0.0479&2.0039&2.9571&71.44\\
 & Loser (L) &0.2781&4.2244&-1.2474&8.0932&1.4323&0.0888&1.6588&2.3910&64.92\\
 & L -- W &0.2416&2.3988&0.2320&1.0899&1.2443&0.0422&1.5481&2.1681&20.36\\
\\[-2ex] 
$p^{(1)}(\tau,r)$& Winner (W) &0.0433&4.3384&-1.2263&6.4733&0.2229&0.0487&2.0007&2.9610&71.20\\
 & Loser (L) &0.2986&4.2776&-1.2293&8.1616&1.5376&0.0891&1.6558&2.3985&65.17\\
 & L -- W &0.2553&2.4203&0.2482&1.3236&1.3147&0.0431&1.5482&2.1747&21.30\\
\\[-2ex] 
$p^{(1)}(\upsilon,R)$& Winner (W) &0.0399&4.3078&-1.2111&6.2480&0.2053&0.0482&1.9968&2.9472&70.77\\
 & Loser (L) &0.3006&4.3159&-1.2610&8.4086&1.5479&0.0892&1.6627&2.4087&65.53\\
 & L -- W &0.2607&2.4439&0.2286&1.4620&1.3427&0.0450&1.5289&2.1522&21.51\\
\\[-2ex] 
$p^{(1)}(\tau,R)$& Winner (W) &0.0481&4.2810&-1.1999&6.4031&0.2479&0.0490&1.9824&2.9382&70.31\\
 & Loser (L) &0.3002&4.3536&-1.2939&8.5965&1.5462&0.0886&1.6676&2.4188&66.07\\
 & L -- W &0.2521&2.4656&0.1705&1.6283&1.2983&0.0422&1.5293&2.1521&23.29\\
\\[-2ex] 
$p^{(2)}(\upsilon,r)$& Winner (W) &0.1214&3.8635&-1.3671&7.6297&0.6251&0.0610&1.8278&2.6834&68.73\\
 & Loser (L) &0.2842&3.9333&-1.6138&11.3616&1.4637&0.0951&1.4908&2.1750&65.07\\
 & L -- W &0.1628&2.1709&0.2235&1.4173&0.8386&0.0281&1.4383&2.0149&21.29\\
\\[-2ex] 
$p^{(2)}(\tau,r)$& Winner (W) &0.1407&3.8520&-1.3704&7.7786&0.7244&0.0624&1.8156&2.6577&68.59\\
 & Loser (L) &0.2891&3.9623&-1.5753&11.1774&1.4889&0.0945&1.4799&2.1615&65.38\\
 & L -- W &0.1484&2.1738&0.2044&1.4014&0.7645&0.0280&1.4775&2.0766&20.54\\
\\[-2ex] 
$p^{(2)}(\upsilon,R)$& Winner (W) &0.1389&3.8274&-1.3849&7.7040&0.7151&0.0624&1.8149&2.6574&68.95\\
 & Loser (L) &0.2875&3.9804&-1.5728&11.0874&1.4807&0.0940&1.5239&2.2234&65.53\\
 & L -- W &0.1487&2.1910&0.2024&1.6035&0.7657&0.0288&1.5139&2.1327&20.77\\
\\[-2ex] 
$p^{(2)}(\tau,R)$& Winner (W) &0.1597&3.8158&-1.4078&7.9932&0.8223&0.0654&1.8028&2.6433&69.02\\
 & Loser (L) &0.2886&4.0066&-1.5608&10.9274&1.4865&0.0932&1.5209&2.2186&65.49\\
 & L -- W &0.1290&2.2188&0.1553&1.6056&0.6642&0.0258&1.5409&2.1789&21.88\\
\\[-2ex] 
$p^{(3)}(1/\sigma,r)$& Winner (W) &0.2722&3.4612&-1.4260&7.9900&1.4019&0.0794&1.5697&2.2775&62.39\\
 & Loser (L) &0.2486&3.8970&-1.3369&8.6249&1.2801&0.0887&1.4988&2.1919&59.58\\
 & L -- W &-0.0237&2.5536&0.3605&2.1385&-0.1218&0.0055&1.5935&2.2586&49.46\\
\\[-2ex] 
$p^{(3)}(1/\sigma,R)$& Winner (W) &0.2871&3.4144&-1.3862&7.6921&1.4783&0.0813&1.5474&2.2447&61.26\\
 & Loser (L) &0.2401&3.9418&-1.3097&8.6519&1.2364&0.0865&1.5137&2.2094&59.65\\
 & L -- W &-0.0470&2.6091&0.3298&2.3515&-0.2419&0.0036&1.6149&2.2948&55.65\\
\hline
\end{tabular}
\label{tbl_daily_summary_stat_risk_phys_momentum_weekly_6_6_contrarian_kr_kp200}
\end{center}
\end{sidewaystable}

	The outperformance of the $p^{(1)}$ and $p^{(2)}$ portfolios is achieved by the stronger reversal in each ranking group. The average returns of the loser groups in those strategies are in the range of 0.278\%--0.301\% comparing with weekly 0.266\% by the benchmark contrarian portfolio. Additionally, the volatility levels of the loser groups are lower than the loser in cumulative return. Meanwhile, the winner baskets in the alternative portfolios underperform its competitive winner basket. While the traditional mean-reversion winner group obtains weekly 0.198\%, the performance of the winners in $p^{(1)}$ is in the range of 0.037\%--0.048\% and the $p^{(2)}$ winners gain weekly 0.121\%--0.160\% with the smaller standard deviations than the other strategies. In the cases of the $p^{(3)}$ criteria, the loser groups are as good as the other loser baskets but the winner groups show the strongest momentum. 
	
	An interesting finding in Table \ref{tbl_daily_summary_stat_risk_phys_momentum_weekly_6_6_contrarian_kr_kp200} is that the outperformance of all the $p^{(1)}$ and $p^{(2)}$ portfolios is achieved by taking low risk. All the strategies in these classes are less riskier in every risk measure such as 95\% VaR, CVaR, and maximum drawdown. The maximum drawdowns are almost 50\% decreased with respect to the benchmark strategy. Additionally, the portfolios in the both categories exhibit much higher Sharpe ratios. In particular, the risk measures of every selection rule in $p^{(1)}$ are at the lowest levels. The risk measures and Sharpe ratios of these alternative portfolios are in the narrow ranges. This fact indicates that the consistent risk management is valid for any choices of price momentum definition. The $p^{(2)}$ strategies also show the same pattern: higher Sharpe ratios and lower risk measures in the narrow ranges. Additionally, the risk measures are slightly lower than any other contrarian portfolios.
	
	The risk profile of each ranking basket is also consistent with the purpose of the ranking group. Every alternative loser is less riskier than the benchmark loser and the winner groups are much exposed to the risk. In particular, the 95\% VaR and CVaR levels of the alternative loser groups are lower than those of the loser in cumulative return. Meanwhile, the winner groups are under greater exposure to the risk than the winner basket in the traditional contrarian portfolio. The same pattern is observed for the Sharpe ratio. The Sharpe ratios of the loser (winner) baskets in $p^{(1)}$ and $p^{(2)}$ portfolios are greater (smaller) than that of cumulative return loser. The $p^{(2)}$ losers (winners) exhibit the slightly better (worse) reward--risk measures than $p^{(1)}$.

\subsection{U.S. equity market: S\&P 500}
\begin{figure}[h!]
	\begin{center}
		\includegraphics[width=8cm]{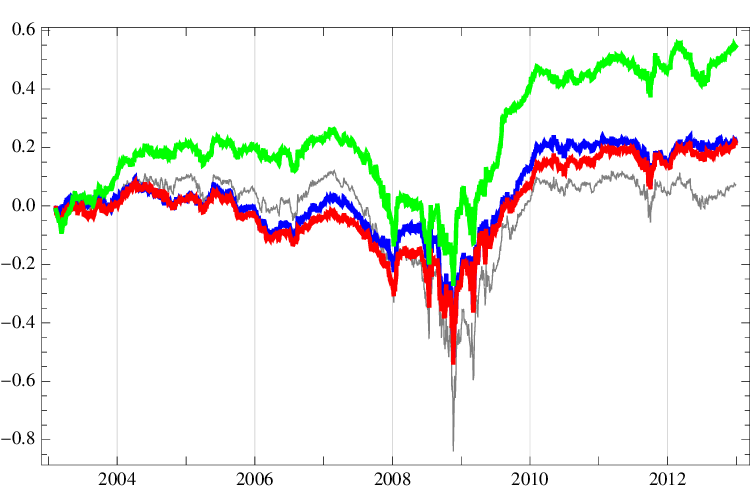}
	\end{center}
		\caption{Cumulative returns for the traditional contrarian (gray), $p^{(1)}(\upsilon,R)$ (blue), $p^{(2)}(\upsilon,R)$ (red), and $p^{(3)}(1/\sigma,R)$ (green) in U.S. S\&P 500. (For interpretation of the references to color in this figure legend, the reader is referred to the web version of this article.)}
		\label{grp_acc_return_phys_momentum_weekly_6_6_contrarian_us_spx}
\end{figure}

	In Table \ref{tbl_daily_summary_stat_risk_phys_momentum_weekly_6_6_contrarian_us_spx}, every alternative momentum strategy outperforms the traditional contrarian strategy. The best portfolios are from the $p^{(3)}$ criteria with weekly 0.107\% and 0.106\%, respectively.  These average returns are almost seven-times greater than the performance of the original mean-reversion portfolio. Additionally, the volatility levels of the portfolio performance are almost 50\% decreased with respect to the benchmark strategy. The $p^{(3)}$ portfolios also exhibit the higher skewness and lower kurtosis. The strategies constructed by $p^{(1)}$ and $p^{(2)}$ also obtain the better performance under the smaller standard deviation. However, the skewness is decreased with respect to the cumulative return strategy. There is no significant improvement in kurtosis. The performance of the alternative portfolios can be found in Fig. \ref{grp_acc_return_phys_momentum_weekly_6_6_contrarian_us_spx}.

	\begin{sidewaystable}
\begin{center}
\caption{Summary statistics and risk measures of weekly 6/6 contrarian portfolios in U.S. S\&P 500}
\scriptsize
\begin{tabular}{l l r r r r r r r r r}
\hline
Criterion & Portfolio & \multicolumn{5}{l}{Summary statistics}  & \multicolumn{4}{l}{Risk measures}\\ \cline{3-7} \cline{8-11} 
 & & Mean & Std. Dev. & Skewness & Kurtosis & Fin. Wealth & Sharpe & $\textrm{VaR}_{95\%}$ & $\textrm{CVaR}_{95\%}$ & MDD \\ 
\hline
$p^{(0)}$ & Winner (W) &0.1545&3.4269&-0.8186&6.8648&0.7970&0.0560&1.3386&1.7016&63.77\\
 & Loser (L) &0.1685&4.6130&-0.1908&11.7318&0.8693&0.0494&1.3727&1.7022&81.06\\
 & L -- W &0.0140&2.7235&0.5559&18.0504&0.0722&-0.0018&0.5749&0.7774&68.14\\
\\[-2ex] 
$p^{(1)}(\upsilon,r)$& Winner (W) &0.1480&4.0676&-0.5063&7.2687&0.7636&0.0507&1.4203&1.8365&71.54\\
 & Loser (L) &0.1807&4.3508&-0.0870&9.3735&0.9326&0.0541&1.4668&1.8387&77.10\\
 & L -- W &0.0327&1.9516&-0.5294&17.1004&0.1690&-0.0013&0.6181&0.8700&42.67\\
\\[-2ex] 
$p^{(1)}(\tau,r)$& Winner (W) &0.1488&3.9203&-0.5343&6.6807&0.7679&0.0507&1.4777&1.9006&70.37\\
 & Loser (L) &0.1877&4.4446&-0.0486&9.7044&0.9684&0.0530&1.4598&1.8298&77.38\\
 & L -- W &0.0388&2.0502&0.1013&14.3828&0.2004&-0.0004&0.6148&0.8658&47.63\\
\\[-2ex] 
$p^{(1)}(\upsilon,R)$& Winner (W) &0.1500&3.8309&-0.5550&6.2724&0.7738&0.0517&1.4125&1.8306&69.49\\
 & Loser (L) &0.1930&4.4955&-0.0530&9.7263&0.9960&0.0522&1.4650&1.8419&78.09\\
 & L -- W &0.0431&2.1074&0.3413&14.9428&0.2221&-0.0002&0.6133&0.8613&51.78\\
\\[-2ex] 
$p^{(1)}(\tau,R)$& Winner (W) &0.1537&3.6880&-0.5759&5.5509&0.7931&0.0518&1.4104&1.8286&67.46\\
 & Loser (L) &0.1937&4.5815&-0.0799&10.3110&0.9997&0.0515&1.4634&1.8409&78.35\\
 & L -- W &0.0401&2.2530&0.4616&20.3251&0.2067&0.0003&0.6102&0.8587&56.16\\
\\[-2ex] 
$p^{(2)}(\upsilon,r)$& Winner (W) &0.1402&3.7924&-0.5654&8.0133&0.7236&0.0519&1.3162&1.6957&71.08\\
 & Loser (L) &0.1786&4.0685&-0.1258&9.3141&0.9217&0.0564&1.4230&1.7917&76.01\\
 & L -- W &0.0384&1.8332&-0.3775&16.0465&0.1981&0.0011&0.5720&0.7977&42.57\\
\\[-2ex] 
$p^{(2)}(\tau,r)$& Winner (W) &0.1435&3.6434&-0.6414&7.2830&0.7404&0.0526&1.3151&1.6962&69.64\\
 & Loser (L) &0.1805&4.1670&-0.1287&9.9820&0.9314&0.0559&1.4263&1.7964&76.92\\
 & L -- W &0.0370&1.9170&0.0761&13.1334&0.1910&0.0017&0.5696&0.7971&49.00\\
\\[-2ex] 
$p^{(2)}(\upsilon,R)$& Winner (W) &0.1416&3.5544&-0.6954&6.9265&0.7308&0.0530&1.3084&1.6894&68.87\\
 & Loser (L) &0.1834&4.2449&-0.1108&10.3969&0.9462&0.0556&1.4250&1.7955&77.25\\
 & L -- W &0.0417&1.9983&0.3715&14.2958&0.2154&0.0047&0.5727&0.7993&51.69\\
\\[-2ex] 
$p^{(2)}(\tau,R)$& Winner (W) &0.1504&3.4075&-0.7225&6.0400&0.7759&0.0544&1.3142&1.7000&67.00\\
 & Loser (L) &0.1877&4.3447&-0.1401&11.1255&0.9687&0.0555&1.4290&1.8006&77.82\\
 & L -- W &0.0374&2.1262&0.4476&19.4093&0.1927&0.0047&0.5662&0.7880&56.10\\
\\[-2ex] 
$p^{(3)}(1/\sigma,r)$& Winner (W) &0.0920&2.8331&-1.0490&8.6160&0.4746&0.0591&1.1677&1.4724&61.18\\
 & Loser (L) &0.1992&3.4731&-0.4319&8.3765&1.0281&0.0635&1.2578&1.5768&70.11\\
 & L -- W &0.1073&1.7319&0.7020&5.0542&0.5535&0.0158&0.4963&0.6816&40.08\\
\\[-2ex] 
$p^{(3)}(1/\sigma,R)$& Winner (W) &0.0968&2.6887&-1.0728&7.7348&0.4995&0.0598&1.1644&1.4695&59.05\\
 & Loser (L) &0.2024&3.5789&-0.3583&8.8179&1.0444&0.0620&1.2602&1.5817&71.17\\
 & L -- W &0.1056&1.8802&0.8292&7.1623&0.5449&0.0145&0.5023&0.6938&45.90\\
\hline
\end{tabular}
\label{tbl_daily_summary_stat_risk_phys_momentum_weekly_6_6_contrarian_us_spx}
\end{center}
\end{sidewaystable}

	In weekly scale, the contrarian strategies exhibit the strong reversal in each ranking group. First of all, all loser groups outperform the traditional contrarian loser group. In particular, the performance of the two $p^{(3)}$ loser baskets is much better than any other alternative and benchmark loser baskets. Additionally, the standard deviations of these $p^{(3)}$ loser groups are much lower than those of the other losers constructed by the $p^{(0)}$, $p^{(1)}$, and $p^{(2)}$ physical momentum. Opposite to the loser groups, the winner groups underperform the winner basket in cumulative return. Although the performance of other winner groups is slightly worse than the short basket in cumulative return portfolio, the winner baskets in the $p^{(3)}$ criteria achieve the lowest average returns.
		
	According to Table  \ref{tbl_daily_summary_stat_risk_phys_momentum_weekly_6_6_contrarian_us_spx}, the $p^{(3)}$ portfolios are less riskier in 95\% VaR, CVaR, and maximum drawdown. The portfolio with inverse volatility and raw return for the physical momentum definition achieves the lowest risk measures. Additionally, its Sharpe ratio is the largest Sharpe ratio among all the alternative portfolios including the benchmark. Another $p^{(3)}$ strategy is ranked the next in every measure to the $p^{(3)}(1/\sigma,r)$ portfolio. Meanwhile, other alternative contrarian portfolios constructed by the physical momentum are riskier in 95\% VaR and CVaR although the maximum drawdowns are improved. Moreover, the Sharpe ratios are better than the benchmark case but much lower than the $p^{(3)}$ portfolios.
	
	The $p^{(3)}$ portfolios are less riskier at the level of each ranking basket. The winner and lower baskets in the $p^{(3)}$ portfolios achieve the lowest 95\% VaR, CVaR, and maximum drawdown among every competitive basket including the traditional contrarian portfolio. Additionally, the Sharpe ratios of the loser groups in the $p^{(3)}$ portfolios are the largest Sharpe ratios. The winner groups of the $p^{(3)}$ portfolios also obtain the highest reward--risk measures. The ranking baskets in the $p^{(2)}$ portfolios are slightly less risker but the winner and loser groups in the  $p^{(1)}$ definitions exhibit the worse risk measures.

\section{Factor analysis}
\label{sec_phys_factor_analysis}
	The intercepts and factor exposures of the Fama--French three-factor analysis on the S\&P 500 results are given in Table \ref{tbl_ff_regression_phys_momentum_weekly_6_6_contrarian_us_spx}. All the intercepts by the physical momentum definitions are greater than the three-factor alpha of the traditional contrarian strategy. In particular, the $p^{(3)}$ portfolios achieve the largest and positive three-factor alphas.
	
	\begin{table}[h!]
\begin{center}
\caption{Fama--French regression of weekly 6/6 contrarian portfolios in U.S. S\&P 500}
\scriptsize
\begin{tabular}{l l r @{} l r @{} l r @{} l r @{} l r}
\hline
Criterion & Portfolio & \multicolumn{9}{l}{Factor loadings} \\ \cline{3-11} 
 & & \multicolumn{2}{c}{$\alpha(\%)$} & \multicolumn{2}{c}{$\beta_{MKT}$} & \multicolumn{2}{c}{$\beta_{SMB}$} & \multicolumn{2}{c}{$\beta_{HML}$} & $R^2$ \\ 
\hline
$p^{(0)}$ &Winner (W)&-0.0494&&1.1237&${}^{**}$&0.2532&${}^{**}$&0.2906&${}^{**}$&0.8744\\ 
&Loser (L)&-0.0965&&1.4406&${}^{**}$&0.1721&${}^{*}$&0.6591&${}^{**}$&0.8445\\ 
&L -- W&-0.0471&&0.3169&${}^{**}$&-0.0810&&0.3685&${}^{**}$&0.1602\\ 
\\[-2ex]$p^{(1)}(\upsilon,r)$&Winner (W)&-0.1003&&1.2949&${}^{**}$&0.3612&${}^{**}$&0.5236&${}^{**}$&0.8936\\ 
&Loser (L)&-0.0762&&1.4045&${}^{**}$&0.2632&${}^{**}$&0.4818&${}^{**}$&0.8704\\ 
&L -- W&0.0241&&0.1096&${}^{**}$&-0.0980&&-0.0418&&0.0172\\ 
\\[-2ex]$p^{(1)}(\tau,r)$&Winner (W)&-0.0901&&1.2575&${}^{**}$&0.3567&${}^{**}$&0.4547&${}^{**}$&0.8911\\ 
&Loser (L)&-0.0744&&1.4219&${}^{**}$&0.2594&${}^{**}$&0.5389&${}^{**}$&0.8680\\ 
&L -- W&0.0157&&0.1644&${}^{**}$&-0.0973&&0.0841&&0.0518\\ 
\\[-2ex]$p^{(1)}(\upsilon,R)$&Winner (W)&-0.0837&&1.2310&${}^{**}$&0.3670&${}^{**}$&0.4165&${}^{**}$&0.8890\\ 
&Loser (L)&-0.0719&&1.4297&${}^{**}$&0.2567&${}^{**}$&0.5782&${}^{**}$&0.8673\\ 
&L -- W&0.0118&&0.1986&${}^{**}$&-0.1102&&0.1617&${}^{*}$&0.0849\\ 
\\[-2ex]$p^{(1)}(\tau,R)$&Winner (W)&-0.0708&&1.1925&${}^{**}$&0.3693&${}^{**}$&0.3448&${}^{**}$&0.8842\\ 
&Loser (L)&-0.0762&&1.4482&${}^{**}$&0.2549&${}^{**}$&0.6252&${}^{**}$&0.8671\\ 
&L -- W&-0.0053&&0.2557&${}^{**}$&-0.1144&&0.2805&${}^{**}$&0.1450\\ 
\\[-2ex]$p^{(2)}(\upsilon,r)$&Winner (W)&-0.0885&&1.2362&${}^{**}$&0.2405&${}^{**}$&0.4664&${}^{**}$&0.9034\\ 
&Loser (L)&-0.0603&&1.3269&${}^{**}$&0.1972&${}^{**}$&0.4441&${}^{**}$&0.8757\\ 
&L -- W&0.0282&&0.0907&${}^{*}$&-0.0433&&-0.0222&&0.0139\\ 
\\[-2ex]$p^{(2)}(\tau,r)$&Winner (W)&-0.0764&&1.2063&${}^{**}$&0.2370&${}^{**}$&0.3831&${}^{**}$&0.9074\\ 
&Loser (L)&-0.0638&&1.3386&${}^{**}$&0.2013&${}^{**}$&0.5146&${}^{**}$&0.8699\\ 
&L -- W&0.0127&&0.1323&${}^{**}$&-0.0357&&0.1315&&0.0513\\ 
\\[-2ex]$p^{(2)}(\upsilon,R)$&Winner (W)&-0.0731&&1.1870&${}^{**}$&0.2396&${}^{**}$&0.3311&${}^{**}$&0.9088\\ 
&Loser (L)&-0.0650&&1.3536&${}^{**}$&0.1942&${}^{**}$&0.5621&${}^{**}$&0.8681\\ 
&L -- W&0.0080&&0.1666&${}^{**}$&-0.0454&&0.2310&${}^{**}$&0.0928\\ 
\\[-2ex]$p^{(2)}(\tau,R)$&Winner (W)&-0.0553&&1.1538&${}^{**}$&0.2384&${}^{**}$&0.2512&${}^{**}$&0.9110\\ 
&Loser (L)&-0.0660&&1.3701&${}^{**}$&0.1897&${}^{**}$&0.6280&${}^{**}$&0.8656\\ 
&L -- W&-0.0107&&0.2163&${}^{**}$&-0.0487&&0.3768&${}^{**}$&0.1676\\ 
\\[-2ex]$p^{(3)}(1/\sigma,r)$&Winner (W)&-0.0721&&1.0048&${}^{**}$&0.0652&&0.0936&${}^{**}$&0.9099\\ 
&Loser (L)&-0.0012&&1.1691&${}^{**}$&0.0816&&0.3046&${}^{**}$&0.8830\\ 
&L -- W&0.0709&&0.1642&${}^{**}$&0.0164&&0.2111&${}^{**}$&0.1188\\ 
\\[-2ex]$p^{(3)}(1/\sigma,R)$&Winner (W)&-0.0582&&0.9689&${}^{**}$&0.0661&${}^{*}$&0.0187&&0.9125\\ 
&Loser (L)&-0.0042&&1.1906&${}^{**}$&0.0865&&0.3578&${}^{**}$&0.8791\\ 
&L -- W&0.0540&&0.2217&${}^{**}$&0.0204&&0.3391&${}^{**}$&0.2082\\ 
\hline
${}^{**}$ 1\% significance & ${}^{*}$ 5\% significance  
\end{tabular}
\label{tbl_ff_regression_phys_momentum_weekly_6_6_contrarian_us_spx}
\end{center}
\end{table}
	
	The market and value factors are statistically significant. Meanwhile, the size factor does not show any significance in any portfolios. The factor loadings on the size factor are positive only in the $p^{(3)}$ portfolios. For any contrarian portfolios in the S\&P 500 universe, the performance of the contrarian portfolios are not explicable with the Fama--French three-factor model.
	
	For each ranking basket, the intercepts of the regression are all negative. The stronger reversal in the ranking baskets is also found in the Fama--French three-factor analysis. The largest loser alphas are achieved by the winner groups in the $p^{(3)}$ portfolios. The alphas of the loser groups are exceptionally smaller than other loser baskets. The worst alphas are obtained by the $p^{(1)}$ portfolios. The three-factor alphas of the $p^{(3)}$ losers are also worse than the cumulative return case. 
	
	The factor exposures of the contrarian portfolios are similar to other strategies except for $p^{(3)}$. Although all the factor loadings for the winner and loser baskets in $p^{(0)},p^{(1)},$ and $p^{(2)}$ are positive and statistically significant, the $p^{(3)}$ strategies exhibit the weaker dependence on the size factor. The Fama--French three-factor model can explain the large parts of the portfolio performance with high $R^2$ values.

\section{Conclusion}	
\label{sec_phys_conclusion}
	In this paper, the various definitions of the physical momentum on equity price are introduced. Using the mapping between the price of a financial instrument and position of a particle in the one-dimensional space, the log return corresponds to the velocity in the log-price space. Up to the higher-order correction terms, the cumulative return is also considered as the velocity. The candidates for the financial mass to define the equity momentum quantitatively are the fractional volume, fractional transaction amount in cash, and the inverse of volatility. These definitions have the plausible origins not only from the viewpoint of physics but also based on financial viewpoint.
	
	With the financial mass and velocity concepts, it is capable of defining the physical momentum in asset price that is called as the price momentum in finance. Measuring the physical momentum for each equity, the contrarian strategies using the physical momentum as a ranking criteria are implemented in the KOSPI 200 and S\&P 500 universes. 
	
	Its performance and reward--risk ratios surpass those of the traditional contrarian strategy in the weekly level. The outperformance of the physical momentum definition is based on the strong mean-reversion in each ranking basket. The winner in physical momentum definition underperforms the winner group of the traditional contrarian portfolio.

	The performance of the physical momentum portfolios is not explained by the Fama--French three-factor model. The intercepts are higher than the cumulative return strategy and the $R^2$ values are much lower.
	
	In future work, the similar test will be conducted in different markets and asset classes. It will be interesting to implement the physical momentum portfolios in different trading strategies such as high frequency.

\section*{Acknowledgement}
	We are thankful to Sungsoo Choi, Jonghyoun Eun, Wonseok Kang, and Svetlozar Rachev for useful discussions.

\end{document}